\documentclass[12pt]{article}
\usepackage{amssymb,amsmath,epsfig}

\begin{document}
\title{\bf Expansion-free Cylindrically Symmetric Models}

\author{M. Sharif \thanks{msharif.math@pu.edu.pk} and Z. Yousaf
\thanks{z.yousaf.math@live.com}\\
Department of Mathematics, University of the Punjab,\\
Quaid-e-Azam Campus, Lahore-54590, Pakistan.}

\date{}

\maketitle
\begin{abstract}
This paper investigates cylindrically symmetric distribution of
an-isotropic fluid under the expansion-free condition, which
requires the existence of vacuum cavity within the fluid
distribution. We have discussed two family of solutions which
further provide two exact models in each family. Some of these
solutions satisfy Darmois junction condition while some show the
presence of thin shell on both boundary surfaces. We also formulate
a relation between the Weyl tensor and energy density.
\end{abstract}

\section{Introduction}

In general relativity, the expansion scalar measures the change of
small volumes of the fluid with respect to time.  An explosion in
the center leads to an overall expansion of the fluid thus making a
cavity surrounding the center. Skripkin \cite{1} was the pioneer who
described the fascinating phenomenon of cavity formation by assuming
non-dissipative isotropic fluid. This fluid was initially at rest
but on explosion at the center, fluid ejects outwards thus forming
the Minkowskian cavity inside. The expansion scalar vanishes for all
such solutions.

Herrera et al. \cite{2} discussed the physical meaning of
expansion-free motion with the help of two different definitions for
the radial velocity of a fluid element. It was found \cite{3} that
the Skripkin model does not satisfy Darmois junction conditions. In
most cases, expansion-free models require pressure anisotropy and
energy density inhomogeneity \cite{6,7}. Some people
\cite{8}-\cite{10} suggested that anisotropy plays a vital role for
understanding the stability of highly compact bodies. Herrera et al.
\cite{11} described the instability of the cavity with the
expansion-free fluid distribution. Sharif and his collaborators
\cite{11a} have also discussed the dynamical instability of
expansion-free gravitational collapse. The expansion-free condition
also helps to explain voids.

The cavity evolution problem requires junction conditions which join
two distinct solutions into one. Here we have two hypersurfaces
delimiting the fluid, external and internal. The former delimits
fluid distribution from cylindrically metric and the later delimits
the Minkowskian cavity. The first set of junction conditions in
general relativity was introduced by Darmois \cite{27}. In a recent
paper, Di Prisco et al. \cite{30} found some exact analytic models
of spherically symmetric spacetime under expansion-free condition
some satisfying the junction conditions.

In this paper, we take the expansion-free cylindrically symmetric
distribution of anisotropic fluid with a vacuum cavity. This paper
is organized as follows. In the next section, we formulate the field
equations and review some basic properties of anisotropic fluid.
Section \textbf{3} provides the relationship between the Weyl tensor
and energy density. In section \textbf{4}, some analytic solutions
are obtained under the expansion-free condition. We summarize the
results in the last section.

\section{Fluid Distribution and the Field Equations}

Consider a cylindrically symmetric distribution of collapsing
fluid bounded by a cylindrical surface $\Sigma^{(e)}$. The
interior region is given by \cite{31}
\begin{equation}\label{1}
ds^2_-=-A^2(t,r)dt^{2}+B^2(t,r)dr^{2}+C^2(t,r){d\phi^{2}}+ dz^2,
\end{equation}
where $-\infty{\leq}t{\leq}\infty,~0\leq{r}\leq\infty,~
0\leq{\phi}\leq{2\pi},~-\infty<{z}<{\infty}$ and the comoving
coordinates are taken inside the hypersurface $\Sigma^{(e)}$. The
energy-momentum tensor $T^-_{\alpha\beta}$
 for a locally anisotropic fluid is \cite{32}
\begin{equation}\label{2}
T^-_{\alpha\beta}=(\mu+P_{r})V_{\alpha} V_{\beta}+P_r
g_{\alpha\beta} + (P_{z}-P_{r})S_{\alpha} S_{\beta}+
(P_\phi-P_{r})K_{\alpha} K_{\beta},
\end{equation}
where $\mu,~P_{r},~P_{\phi},~P_{z},~V^{\alpha},~K^{\alpha}$ and
$S_{\alpha}$ are the energy density, the principal stresses, four
velocity and four-vectors, respectively, satisfying
\begin{equation*}
V^{\alpha}V_{\alpha}=-1,\quad
S^{\alpha}S_{\alpha}=K^{\alpha}K_{\alpha}=1,\quad
S^{\alpha}K_{\alpha}=V^{\alpha}K_{\alpha}=V^{\alpha}S_{\alpha}=0,
\end{equation*}
with
\begin{equation*}\label{3}
V_{\alpha}=-A\delta^{0}_{\alpha}, \quad
K_{\alpha}=C{\delta}^{2}_{\alpha},\quad
S_{\alpha}={\delta}^{3}_{\alpha}.
\end{equation*}
The four acceleration and its non-vanishing components are
\begin{equation*}\label{4}
a_\alpha=V_{\alpha;\beta}V^\beta,\quad a_{1}=\frac{A'}{A},\quad
a^2=a^\alpha a_{\alpha}=\left(\frac{A'}{AB}\right)^2.
\end{equation*}
The shear tensor is defined as
\begin{equation*}\label{5}
\sigma_{\alpha\beta}=V_{(\alpha;b)}+a_{(\alpha} V_{\beta)}
-\frac{1}{3}\Theta(g_{\alpha\beta}+V_\alpha V_\beta),
\end{equation*}
with non-zero components
\begin{equation*}\label{7}
\sigma_{11}=\frac{B^{2}}{3A}\left(\frac{2\dot{B}}{B}-\frac{\dot{C}}{C}\right),\quad
\sigma_{22}=\frac{-C^{2}}{3A}\left(\frac{\dot{B}}{B}-\frac{2\dot{C}}{C}\right),\quad
\sigma_{33}=\frac{-1}{3A}\left(\frac{\dot{B}}{B}+\frac{\dot{C}}{C}\right),
\end{equation*}
where dot and prime stand for $t$ and $r$ differentiation,
respectively. The expansion scalar is
\begin{equation}\label{6}
\Theta=V^{\alpha}_{~;\alpha}=\frac{1}{A}\left(\frac{\dot{B}}{B}
+\frac{\dot{C}}{C}\right).
\end{equation}

The field equations for the interior spacetime (\ref{1}) are
\begin{eqnarray}\label{8}
{\kappa}{\mu}A^{2}&=&\left(\frac{A}{B}\right)^2
\left(\frac{B'C'}{BC}-\frac{C''}{C}\right)+\frac{\dot{B}\dot{C}}{BC},\\\label{9}
0&=&\frac{\dot{C'}}{C}-\frac{\dot{C}A'}{CA}-\frac{\dot{B}C'}{BC},\\\label{10}
{\kappa}P_{r}B^{2}&=&\left(\frac{B}{A}\right)^2
\left(\frac{\dot{A}\dot{C}}{AC}-\frac{\ddot{C}}{C}\right)+\frac{A'C'}{AC},
\end{eqnarray}
\begin{eqnarray}\label{11}
{\kappa}P_{\phi}&=&\left(\frac{1}{AB}\right)
\left(\frac{A''}{B}-\frac{\ddot{B}}{A}+\frac{\dot{A}\dot{B}}{A^2}
-\frac{A'B'}{B^2}\right),\\\nonumber
{\kappa}P_{z}&=&\frac{A''}{AB^2}-\frac{\ddot{B}}{A^2B}+\frac{\dot{A}\dot{B}}{A^3B}
-\frac{A'B'}{AB^3}+\frac{\dot{A}\dot{C}}{A^3C}-\frac{\ddot{C}}{A^2C}\\\label{12}
&-&\frac{B'C'}{B^3C}+\frac{C''}{B^2C}+\frac{A'C'}{AB^2C}-\frac{\dot{B}\dot{C}}{A^2BC}.
\end{eqnarray}
Thorne defined C-energy as \cite{33}
\begin{equation}\label{13}
E(t,r)=\frac{1}{8}(1-l^{-2}{\nabla}^{\alpha}\tilde{r}{\nabla}_{\alpha}\tilde{r}),
\end{equation}
where $\rho,~l$ and $\tilde{r}$ are the circumference radius,
specific length and areal radius with the following relations
\begin{equation*}\label{14}
\rho^2={\xi_{(1)a}}{\xi^a_{(1)}},\quad
l^2={\xi_{(2)a}}{\xi^a_{(2)}},\quad \tilde{r}=\rho{l}.
\end{equation*}
Here $\xi_{(1)}=\frac{\partial}{\partial{\phi}},\quad
\xi_{(2)}=\frac{\partial}{\partial{z}}$ are the Killing vectors and
$E(t,r)$ represents the gravitational energy per unit specific
length of the cylinder. The specific energy of the cylinder is given
by $\tilde{E}=El$. However, in view of Eq.(\ref{1}), the specific
length, $l^2$, turns out to be $1$. Therefore, the specific energy
of the cylinder in the interior region is written as
\begin{equation}\label{15}
\tilde{E}(t,r)=E(t,r)=\frac{1}{8}\left[1+\left(\frac{\dot{C}}{A}\right)^2
-\left(\frac{C'}{B}\right)^2\right].
\end{equation}
Differentiating Eq.(\ref{15}) with respect to $t$ and $r$, we get
\begin{equation}\label{16}
\dot{E}=-2{\pi}P_{r}C\dot{C},\quad E'=2\pi{\mu}C'C,
\end{equation}
which yields
\begin{equation}\label{17}
\dot{\mu}C'+{P_r}'\dot{C}+(P_r+\mu)\dot{(C'}+C'\frac{\dot{C}}{C})=0.
\end{equation}
Integrating the second of Eq.(\ref{16}), it follows that
\begin{equation}\label{17a}
E=2\pi\int^r_{0}{\mu}C'Cdr.
\end{equation}
Further, integration yields
\begin{equation}\label{18}
\frac{E}{C^{2}}=\pi\mu-\frac{\pi}{C^2}\int^r_{0}{\mu'}{C^2}dr,
\end{equation}
where we have used $E(t,0)=0=C(t,0)$. The velocity of the
collapsing fluid is $U=\frac{\dot{C}}{A}$ for which Eq.(\ref{15})
leads to
\begin{equation}\label{20}
\breve{E}\equiv\frac{C'}{B}=\left[1+U^{2}-8E\right]^{1/2}.
\end{equation}

Now the electric and magnetic parts of the Weyl tensor are
\begin{equation}\label{21}
\hat{E}_{\alpha\beta}=C_{\alpha\mu\beta\nu}V^{\mu}V^{\nu},\quad
H_{\alpha\beta}=\tilde{C}_{\alpha\gamma\beta\delta}V^{\gamma}V^{\delta}=
\frac{1}{2}\epsilon_{\alpha\gamma\epsilon\delta}
C^{\epsilon\delta}_{~~\beta\rho}V^{\gamma}V^{\rho},
\end{equation}
where
$\epsilon_{\alpha\beta\gamma\delta}\equiv\sqrt{-g}\eta_{\alpha\beta\gamma\delta}$.
The non-vanishing components of the Weyl tensor are
\begin{eqnarray}\nonumber
\hat{E}_{11}&=&\frac{B^{2}}{6A^{2}}\left(\frac{\ddot{C}}{C}
-\frac{\ddot{2B}}{B}-\frac{\dot{A}\dot{C}}{AC}-
\frac{2\dot{A}\dot{B}}{AB}+\frac{\dot{B}\dot{C}}{BC}\right)\\\nonumber
&+&\frac{B^{2}}{6}\left(\frac{2A''}{A}-\frac{C''}{C}-\frac{2A'B'}{AB}
+\frac{B'C'}{BC}-\frac{A'C'}{AC}\right),\\\nonumber
\hat{E}_{22}&=&-\frac{2C^{2}}{6A^{2}}\left(\frac{\ddot{C}}{C}
-\frac{\ddot{B}}{2B}-\frac{\dot{A}\dot{C}}{AC}
-\frac{\dot{A}\dot{B}}{2AB}-\frac{\dot{B}\dot{C}}{2BC}\right)\\\nonumber
&+&\frac{C^{2}}{6B^{2}}\left(\frac{A''}{A}+\frac{C''}{C}
-\frac{2A'C'}{AC}-\frac{B'C'}{BC}-\frac{A'B'}{AB}\right),\\\nonumber
\hat{E}_{33}&=&\frac{1}{6A^{2}}\left(\frac{\dot{A}\dot{B}}{AB}
-\frac{\ddot{C}}{C}-\frac{\ddot{B}}{B}
+\frac{\dot{A}\dot{C}}{AC}+\frac{\dot{2B}\dot{C}}{BC}\right)\\\nonumber
&+&\frac{1}{6B^{2}}\left(\frac{A''}{A}-
\frac{2C''}{C}+\frac{A'C'}{AC}+\frac{2B'C'}{BC}-\frac{A'B'}{AB}\right),\\\nonumber
H_{23}&=&H_{32}=\frac{1}{2A^{2}B^{2}C^{2}}\left(C\dot{C'}
-\frac{C\dot{C}A'}{A}-\frac{C\dot{B}A'}{B}\right).
\end{eqnarray}
The conservation of energy-momentum tensor gives
\begin{equation}\label{22}
\dot{\mu}+A{\mu}\Theta+\frac{\dot{B}}{B}{P_{r}}+\frac{\dot{C}}{C}P_{\phi}=0,
\end{equation}
\begin{equation}\label{23}
{P_{r}}'+(\mu+P_r)\frac{A'}{A}+(P_{r}-P_{\phi})\frac{C'}{C}=0.
\end{equation}

For the junction conditions, we take the exterior spacetime as the
cylindrically symmetric metric \cite{34}
\begin{equation}\label{24}
ds^2_+=-\left(-\frac{2M}{R}\right)d\nu^2
-2d{R}d{\nu}+R^2(d\theta^2+{\alpha}^{2}dz^2),
\end{equation}
where $M$ and $\nu$ are the total mass and retarded time,
respectively. Here $\alpha^2=-\frac{\Lambda}{3}$, where $\Lambda$ is
a cosmological constant. For smooth matching of the interior and
exterior regions, Darmois conditions \cite{27} lead to
\begin{eqnarray}\label{41}
&&E-M\overset{\Sigma^{(e)}}=\frac{1}{8},\quad
P_{r}\overset{\Sigma^{(e)}}=0.
\end{eqnarray}
Here $\overset{\Sigma^{(e)}}=$ indicates that the quantities are
evaluated at external hypersurface. This equation shows that the
difference between two masses is equal to $\frac{1}{8}$ as shown
in the adiabatic case \cite{35}. This is due to the least
unsatisfactory definition of the C-energy \cite{33}.

Notice that the expansion-free models require two hypersurfaces, one
is the boundary between the fluid distribution and the external
cylindrically symmetric solution and other separating the central
Minkowskian cavity from the fluid \cite{30}. Taking $\Sigma^{(i)}$
as the boundary surface of that internal vacuum cavity and the fluid
distribution, then matching of the Minkowski spacetime within the
cavity to the fluid gives
\begin{equation}\label{42}
E(t,r)\overset{\Sigma^{(i)}}{=}0,\quad
P_{r}\overset{\Sigma^{(i)}}{=}0.
\end{equation}

\section{The Weyl Tensor and Matter Variables}

In this section, we obtain a relation between the Weyl tensor and
matter variables. The Weyl scalar $\mathcal{C}$ in terms of the
Kretchman scalar, $\mathcal{R}$, is
\begin{equation}\label{43}
\mathcal{C}^2=\mathcal{R}-2R^{\alpha\beta}R_{\alpha\beta}+\frac{1}{3}R^2,
\end{equation}
where the Kretchman scalar
$\mathcal{R}=R^{\alpha\beta\gamma\delta}R_{\alpha\beta\gamma\delta}$
yields
\begin{eqnarray}\label{44}
\mathcal{R}&=&4\left\{\frac{1}{(AB)^4}(R^{0101})^2+\frac{1}{(BC)^4}(R^{1212})^2
+\frac{1}{(AC)^4}(R^{0202})^2\right.\nonumber\\
&-&\left.\frac{1}{2A^2B^2C^4}(R^{1202})^2\right\}.
\end{eqnarray}
The non-zero components of the Riemann tensor in terms of the
Einstein tensor can be written as
\begin{eqnarray*}\label{45}
R^{0101}&=&\frac{1}{(ABC)^{2}}G_{22},\quad
R^{0202}=\frac{1}{(ABC)^{2}}G_{11},\\\label{46}
R^{0212}&=&\frac{1}{(ABC)^{2}}G_{01},\quad
R^{1212}=\frac{1}{(ABC)^{2}}G_{00}.
\end{eqnarray*}
Substituting these values in Eq.(\ref{44}), we obtain
\begin{equation}\label{47}
\mathcal{R}=4\left\{\frac{G_{00}^2}{A^4}+\frac{G_{11}^2}{B^4}+\frac{G_{22}^2}{C^4}
-\frac{2G_{01}^2}{(AB)^2}\right\}.
\end{equation}
The remaining terms of the Weyl scalar are
\begin{eqnarray*}\label{48}
R_{00}&=&A^{2}\left(\frac{G_{11}}{B^{2}}+\frac{G_{22}}{C^{2}}\right),\quad
R_{01}=G_{01},\\\label{49}
R_{22}&=&C^{2}\left(\frac{G_{00}}{A^{2}}-\frac{G_{11}}{B^{2}}\right),\quad
R_{11}=B^{2}\left(\frac{G_{00}}{A^{2}}-\frac{G_{22}}{C^{2}}\right),\\\label{50}
R&=&2\left(\frac{G_{00}}{A^{2}}-\frac{G_{11}}{B^{2}}+\frac{G_{22}}{C^{2}}\right),\\\nonumber
R^{\alpha\beta}R_{\alpha\beta}&=&2\left\{\frac{G_{00}^2}{A^{4}}+\frac{G_{11}^2}{B^{4}}
+\frac{G_{22}^2}{C^{4}}-\frac{G_{01}^2}{(AB)^{2}}-\frac{G_{00}G_{11}}{(AB)^{2}}\right.\\\label{51}
&+&\left.\frac{G_{11}G_{22}}{(BC)^{2}}-\frac{G_{00}G_{22}}{(AC)^{2}}\right\}.
\end{eqnarray*}
Using the preceding equations, the Weyl scalar (\ref{43}) takes the
form
\begin{eqnarray}\nonumber
\mathcal{C}^{2}&=&\frac{4}{3}\left[\frac{G_{00}^2}{A^4}
+\frac{G_{11}^2}{B^4}+\frac{G_{22}^2}{C^4}
+\frac{G_{00}G_{11}}{(AB)^2}-\frac{G_{11}G_{22}}{(BC)^2}+\frac{G_{22}G_{00}}{(AC)^2}\right].
\end{eqnarray}
Equations (\ref{8}), (\ref{10}) and (\ref{11}) yield
\begin{eqnarray}\label{52a}
\frac{\mathcal{C}\sqrt{3}}{2}&=&\left[\left[\kappa(\mu+P_{r}-P_{\phi})\right]^{2}
-\kappa^{2}{\mu}({P_{r}}-3P_{\phi})+\kappa^{2}P_{r}P_{\phi})\right]^{1/2}.
\end{eqnarray}
When we make use of Eq.(\ref{18}), it follows that
\begin{eqnarray}\nonumber
\frac{\mathcal{C}\sqrt{3}}{2}&=&\left[\left\{\frac{{\kappa}E}{{\pi}C^2}
+\frac{{\kappa}}{C^2}\int^r_{0}{\mu}'{C^2}dr+\kappa(P_{r}-P_{\phi})\right\}^{2}
+\kappa^{2}P_{r}P_{\phi}\right.\\\label{52}
&-&\left.\kappa^{2}({P_{r}}-3P_{\phi})\left({\frac{E}{{\pi}C^2}
+\frac{1}{C^2}\int^r_{0}{\mu}'{C^2}dr}\right)\right]^{1/2},
\end{eqnarray}
which gives the relationship between the Weyl tensor and the fluid
prperties like energy density inhomogeneity and anisotropy of the
pressure.

For pure dust $P_{r}=P_{\phi}=P_{z}=0$, it follows that
\begin{eqnarray}\label{52b}
\frac{\mathcal{C}\sqrt{3}{C^2}}{2}&=&\frac{{\kappa}E}{\pi}
+{\kappa}\int^r_{0}{\mu}'{C^2}dr.
\end{eqnarray}
Using Eq.(\ref{17a}) and differentiating Eq.(\ref{52a}) with
respect to $r$, we have
\begin{eqnarray}\nonumber
\left[\frac{\mathcal{C}\sqrt{3}}{2}\right]'&=&\kappa{\mu'},
\end{eqnarray}
which yields $\mu'=0$ if and only if $\mathcal{C}=constant$. Thus
it is also concluded that if the energy density is homogeneous,
the metric is conformally flat and vice versa as in \cite{28}.

\section{Exact Analytical Models}

Here we use the expansion-free condition to investigate some exact
analytical models. We also check the validity of junction conditions
for the resulting models. The expansion-free condition, $\Theta=0$,
leads to
\begin{equation}\label{53}
B=\frac{\gamma}{C},
\end{equation}
where $\gamma$ is an arbitrary function of $r$. Without loss of
generality, we assume $\gamma=1$. Using this value in Eq.(\ref{9}),
it follows that
\begin{equation}\label{54}
A=\frac{C\dot{C}}{\xi},
\end{equation}
where $\xi$ is an arbitrary function of $t$. The physical
variables $\mu,~P_r$ and $\Pi$ can be written in terms of $C$ and
$E$ as
\begin{equation}\label{55}
2\pi\mu=\frac{E'}{CC'},\quad
2\pi{P_{r}}=-\frac{\dot{E}}{C\dot{C}},
\quad\Pi=(P_{r}-P_{\phi})=\frac{C\dot{\mu}}{\dot{C}}.
\end{equation}
Also, using Eqs.(\ref{15}), (\ref{53}) and (\ref{54}), we can
write
\begin{equation}\label{56}
E=\frac{1}{8}\left(\frac{\xi^{2}}{C^{2}}-C^{2}C'^{2}+1\right).
\end{equation}
We see that the metric coefficients $A$ and $B$ are now expressed in
terms of $C$. In the following, we obtain some exact analytical
models.

\subsection{Solution I}

For the sake of analytical models, we assume $E(t,r)$ of the form
\cite{36}
\begin{equation}\label{57}
2E(t,r)=\frac{1}{3}kC^{3}+\frac{1}{6}lC^{6}+\frac{1}{4},
\end{equation}
where $j,~k$ and $l$ are arbitrary functions of $t$. Using this
value in Eq.(\ref{56}), the energy density becomes
\begin{equation}\label{58}
4\pi\mu=kC+lC^{4}.
\end{equation}
Using Eqs.(\ref{55}) and (\ref{58}), we obtain a differential
equation in terms of $C$
\begin{equation}\label{59}
kC^{2}+lC^{5}+\frac{{\xi}^{2}}{2C^{3}}+\frac{CC'^{2}}{2}+\frac{C^{2}C''}{2}=0,
\end{equation}
which along with Eqs.(\ref{56}) and (\ref{57}) leads to
\begin{equation*}\label{60}
\frac{10}{3}kC^{2}+\frac{8}{3}lC^{5}+3CC'^{2}+C^{2}C''=0.
\end{equation*}
For simplicity, we substitute $C^{2}\equiv{Z}$ so that
\begin{equation*}\label{61}
aZ+bZ^{\frac{5}{2}}+\frac{Z'^{2}}{Z^{\frac{1}{2}}}+Z''Z^{\frac{1}{2}}=0,
\end{equation*}
where $a(t)\equiv\frac{20}{3}k,~b(t)\equiv\frac{16}{3}l$.
Integrating this equation with respect to $Z$, it follows that
\begin{equation}\label{62}
Z'^2=-\frac{4}{7}aZ^{\frac{3}{2}}-\frac{2b}{5}Z^{3}.
\end{equation}
The solution of this equation does not exist explicitly in terms of
$Z$. However, we find some solutions by imposing the following
constraints.

\subsubsection*{Case (i)}

Here, we take $a\neq0$ and $b=0$. Consequently Eq.(\ref{62}) gives
\begin{equation*}\label{63}
Z=\left(\frac{a}{28}\right)^{2}\left(r+\zeta\right)^{4},
\end{equation*}
which can be written as
\begin{equation}\label{64}
C=\left(\frac{5k}{21}\right)\left(r+\zeta\right)^{2},
\end{equation}
where $\zeta(t)$ is an arbitrary function. The corresponding
physical variables turn out to be
\begin{eqnarray}\label{64a}
4\pi\mu&=&\left(\frac{5k^2}{21}\right)\left(r+\zeta\right)^{2},
\\\label{64b}
4\pi{P_r}&=&-5k^2\left(r+\zeta\right)^{2}\left[\frac{\dot{k}
(r+\zeta)}{63\left\{\left(r+\zeta\right)\dot{k}+2k\dot{\zeta}\right\}}
+\frac{1}{21}\right],\\\label{64c}
4\pi{P_{\phi}}&=&-10k^2\left(r+\zeta\right)^{2}\left[\frac{2\dot{k}
(r+\zeta)}{63\left\{\left(r+\zeta\right)\dot{k}+2k\dot{\zeta}\right\}}
+\frac{1}{21}\right], \\\nonumber
8\pi{P_z}&=&\frac{C}{\dot{C}}\left(\dot{C''}C+4\dot{C'}C'+2C''C
+\frac{C'^2\dot{C}}{C}\right)\\\label{65}
&+&\frac{1}{C^3}\left[\frac{\xi\dot{\xi}}{\dot{C}}
+\frac{\dot{C}^2}{C}+\ddot{C}
-\dot{C}-\frac{\ddot{C}\xi^2}{\dot{C}^2}\right].
\end{eqnarray}
When we use Darmois junction conditions, we obtain three independent
equations with two functions $k(t)$ and $\zeta(t)$ which can be
satisfied by any convenient choice of one of these functions.
However, this does not lead to interesting solutions.

\subsubsection*{Case (ii)}

In this case, we take $a=0$ and $b\neq0$. The assumption $a=0$ gives
$k=0$ and hence Eq.(\ref{62}) yields
\begin{equation*}\label{66}
Z=\left(\frac{10}{b}\right)\left(r+\zeta\right)^{-2},
\end{equation*}
or
\begin{equation*}\label{67}
C=\left(r+\zeta\right)^{-1}\left(\frac{15}{8l}\right)^{\frac{1}{2}}.
\end{equation*}
Consequently, the quantities $\mu,~P_r,~P_\phi$ and $P_z$ become
\begin{eqnarray}\nonumber
4\pi\mu&=&\left(\frac{225}{64l^{4}}\right)\left(r+\zeta\right)^{-4},\\\nonumber
4\pi{P_r}&=&\frac{-75}{64l^2\left(r+\zeta\right)^{4}}\left[3l-\frac{\dot{l}l
(r+\zeta)}{\left\{\left(r+\zeta\right)\dot{l}+2l\dot{\zeta}\right\}}\right],\\\nonumber
4\pi{P_\phi}&=&\frac{-75}{64l^2\left(r+\zeta\right)^{4}}\left[15l-\frac{7\dot{l}l
(r+\zeta)}{\left\{\left(r+\zeta\right)\dot{l}+2l\dot{\zeta}\right\}}\right],\\\nonumber
8\pi{P_z}&=&\frac{C}{\dot{C}}\left(\dot{C''}C+4\dot{C'}C'+2C''C
+\frac{C'^2\dot{C}}{C}\right)\nonumber\\
&+&\frac{1}{C^3}\left(\frac{\xi\dot{\xi}}{\dot{C}}\right.
+\left.\frac{\dot{C}^2}{C}+\ddot{C}
-\dot{C}-\frac{\ddot{C}\xi^2}{\dot{C}^2}\right).\label{65}
\end{eqnarray}
On $r=r_{i}$, this subfamily of solution does not satisfy Darmois
conditions.

\subsection{Solution II}

The second family of solution is obtained by assuming
$P_{r}=P_z=0$. It follows from Eq.(\ref{17}) that
$\mu=\frac{d_{1}(r)}{CC'}$ which leads to
\begin{eqnarray}\label{69}
C^{2}=2\int\frac{d_{1}(r)}{\mu}dr+d_{2}(t),
\end{eqnarray}
where $d_{1}$ and $d_{2}$ are integration functions. Consequently,
the first of Eq.(\ref{55}) yields $d_{1}=\frac{E'}{2\pi}$. Under the
expansion-free condition, Eq.(\ref{22}) yields
\begin{equation}\label{71}
P_{\phi}=-\frac{\dot{\mu}C}{\dot{C}}.
\end{equation}
We take the following equation of state to obtain different
physical models.

\subsubsection*{Case (i)}

In this case, we consider $P_{\phi}=\alpha\mu$, where $\alpha$ is a
constant. Using this result in Eq.(\ref{23}), we have
\begin{equation}\label{73}
\dot{C}={f(t)}{C^{\alpha-1}},\quad C'={g(r)}{C^{\alpha-1}},
\end{equation}
or
\begin{equation}\label{74}
C^{2-\alpha}=\psi(t) + \chi(r),
\end{equation}
with $\psi(t)=({2-\alpha})\int{f(t)}dt$ and
$\chi(r)=({2-\alpha})\int{g(r)}dr$. Here $g(r)$ and $f(t)$ are
arbitrary functions. Without loss of generality we can choose
$\xi(t) = f(t)$, then from Eqs.(\ref{54}) and (\ref{73}), we have
$A=C^{\alpha}$. Using the constraints $P_{r}=0$ and
$P_{\phi}={\alpha}{\mu}$ in Eq.(\ref{16}), it implies that $E=E(r)$.
Thus using Eq.(\ref{53}) along with $U=\frac{\dot{C}}{A}$,
Eq.(\ref{15}) becomes
\begin{equation}\label{77}
E(r)=\frac{1}{8}\left(\frac{\dot{C}^{2}}{C^{2\alpha}}-g^{2}(r)C^{2\alpha}+1\right).
\end{equation}

Next, we evaluate this equation on the hypersurfaces, i.e.,
$r=r_{e}$ and $r=r_{i}$
\begin{equation}\label{78}
\dot{C^{2}}\overset{\Sigma^{(i)}}{=}C^{2\alpha}\left(g^{2}C^{2\alpha}-1\right),\quad
\dot{C^{2}}\overset{\Sigma^{(e)}}{=}C^{2\alpha}\left(8M+g^{2}C^{2\alpha}\right).
\end{equation}
From here we can easily evaluate $C$ for an arbitrary value of
$\alpha$. For $\alpha=1/2$, it follows
\begin{equation}\label{80}
C\overset{\Sigma^{(i)}}{=}\left[g^{2}\cos^{2}(t+t_{0})\right]^{-1}.
\end{equation}
Equation(\ref{74}) yields
\begin{equation}\label{81}
\psi(t)\overset{\Sigma^{(i)}}{=}\left[g^{2}\cos^{2}(t+t_{0})\right]^{-3/2}-\chi.
\end{equation}
Thus the time dependence of all variables is fully determined. Now
the radial dependence ($d_1$ or $\chi$) can be obtained from the
initial data.
\begin{eqnarray}\label{82}
{C_{i}}^{2\alpha}(0)-{C_{e}}^{2\alpha}(0)&=&\frac{1}{2g^{2}}\left[1+\left(1+
4g^{2}\dot{C_{i}^2(0)}\right)^{1/2}\right]\nonumber\\
&+&\frac{8M}{2g^{2}}-\frac{\sqrt{4M^{2}+4\dot{C_{e}^{2}}(0)}}{g},
\end{eqnarray}
where ${C_{i}}^{2\alpha}(0)$ and ${C_{i}}^{2\alpha}(0)$ are
calculated from Eq.(\ref{78}) at $t=0$. The differences
${C_{i}}^{2\alpha}(0)-{C_{e}}^{2\alpha}(0)$ and
${C_{i}}^{2\alpha}(0)-{C_{e}}^{2\alpha}(0)$ will be zero for
$\alpha=0$ and negative for any value of $t$ ( if $\alpha>0$)
respectively. Also, from Eqs.(\ref{54}) and $U=\frac{\dot{C}}{A}$,
we obtain
\begin{eqnarray}\label{83}
&&U_{i} = \frac{f}{C_{i}}  =
\frac{\dot{C_{i}}}{C_{i}^{\alpha}},\quad U_{e} = \frac{f}{C_{e}} =
\frac{\dot{C_{e}}}{C_{e}^{\alpha}}.
\end{eqnarray}
As $\alpha<1$ and $C_{e}>C_{i}$, therefore we get
\begin{equation}\label{84}
\dot{C_{i}}>\dot{C _{e}}.
\end{equation}
Thus from Eqs.(\ref{82}) and (\ref{84}), we obtain
${C_{i}}^{2\alpha}>{C_{e}}^{2\alpha}$, which has no physical
significance as such models require the presence of thin shells at
both boundaries.

\subsubsection*{Case (ii)}

Here we assume that energy density is separable so that
\begin{equation}\label{85}
{\mu}={\mu_{0}(t)}d_{1}/r^{2}.
\end{equation}
Consequently, Eqs.(\ref{69}) and (\ref{71}) give
\begin{eqnarray}\label{86}
C=\left(\frac{2r^{3}}{3\mu_{0}}+d_{2}(t)\right)^{\frac{1}{2}},\quad
P_{\phi}=-2\dot{\mu}{\frac{\frac{2r^{3}}{3\mu_{0}}+
d_{2}}{\frac{-r^{3}\dot{\mu_{0}}}{\mu_{0}}+\dot{d_2}}}.
\end{eqnarray}
Also, Eq.(\ref{56}) can be written as
\begin{equation}\label{87}
E=\frac{1}{8}\left[1+\frac{{\xi}^{2}}{C^{2}}-\frac{r^{4}}{{\mu_{0}}^{2}}\right].
\end{equation}
Evaluating the above equation by using Eq.(\ref{41}), we have
\begin{equation}\label{88}
C_{i}^2={\xi}^{2}\left(\frac{r_{i}^{4}}{{\mu_{0}}^{2}}-1\right)^{-1},
\end{equation}
implying that ${r_{i}^{4}}>{\mu_{0}}^{2}$, for all $t$. Using
$U=\frac{\dot{C}}{A}$ and Eq.(\ref{54}) in (\ref{88}), it follows
that
\begin{equation}\label{89}
U_{i}^2=\frac{r_{i}^{4}}{{\mu_{0}}^{2}}-1.
\end{equation}
Thus, from the absence of superluminal velocities $(U<1)$ and from
Eq.(\ref{89}), we have to impose
\begin{equation}\label{90}
r_{i}^2<\sqrt{2}\mu_{0}.
\end{equation}
The relation between $U_{i}$ and $U_{e}$ can be found from
Eq.(\ref{83}) as
\begin{equation}\label{91}
U_{e}=U_{i}\left(\frac{C_{i}}{C_{e}}\right).
\end{equation}
This shows that the inner surface moves faster than the outer one as
$C_{e}>C_{i}$. Further, we take ${\xi}/{C}=\dot{C}$, then
\begin{eqnarray}\label{92}
&&A_i=1,\quad \dot{C_{i}}=U_{i},\quad
\xi={U_{i}}{C_{i}}={U_{e}}{C_{e}}.
\end{eqnarray}
Using Eqs.(\ref{41}), (\ref{87}) and (\ref{92}), we obtain
\begin{equation}\label{93}
C_{e}^{2}={\xi}^{2}\left(8M+\frac{r_{e}^{4}}{{\mu_{0}^{2}}}\right)^{-1},
\end{equation}
which yields
\begin{equation}\label{94}
8M+\frac{r_e^{4}}{{\mu_{0}}^{2}}=U_{e}^{2}.
\end{equation}
Thus, from the absence of superluminal velocities $(U<1)$ and from
Eq.(\ref{94}), we require $r_{e}^{2}<{\mu}_{0}$, which is not the
same as imposed on $r_{i}$ due to the least unsatisfactory
definition of C-Energy. We can find the time dependence of $C_{e}$
if $\mu_{0}=\mu_{0}(t)$ which implies time dependence of all
variables.

\section{Conclusions}

We have found some exact analytical models with expansion-free
condition, total of four solutions, out of which the last two models
satisfy Darmois junction conditions on boundary surfaces. Such
solutions may describe the possible importance of expansion-free
condition to model situation where cavities are expecting to appear.
Also, some of these models may contain essential features of a
realistic situations. It is interesting to note that expansion-free
condition might be helpful for the description of voids. During
expansion, matter of the voids streams out which decreases the
density of voids from inside. They are usually described as a vacuum
cavity around the center. Relaxing Darmois conditions on both
boundary surfaces enlarges the families of possible solutions that
may show their importance in the study of voids with thin wall
approximation \cite{45}. We see from Eq.(\ref{22}) that energy
density changes with time for $\Theta=0$. The results of this paper
demonstrate that expansion-free condition helps a lot in the
modeling of the cavity evolution.

\end{document}